\documentclass{aa}

\usepackage{graphicx}
\usepackage{txfonts}
\usepackage[]{hyperref}
\usepackage{todonotes}
\usepackage{gensymb}

\usepackage{tikz}
\usepackage{booktabs}

\DeclareFontFamily{OT1}{pzc}{}
\DeclareFontShape{OT1}{pzc}{m}{it}{<-> s * [1.10] pzcmi7t}{}
\DeclareMathAlphabet{\mathpzc}{OT1}{pzc}{m}{it}
\DeclareMathOperator{\dtn}{\mathpzc{dtn}}
\newcommand{\DtN}{\mathpzc{DtN}}
\newcommand{\dDtN}{\operatorname{DtN}}

\newcommand{\dmassr}{B}   
\newcommand{\dstiffr}{A}

\newcommand{\ii}{\mathrm{i}}
\newcommand{\dd}{\mathrm{d}}
\newcommand{\kk}{\mathrm{k}}

\newcommand{\sh}{\underline{s}}
\newcommand{\psih}{\underline{\psi}}

\newcommand{\dofs}{DOFs}
\newcommand{\FEshape}{\psi}
\newcommand{\massFEM}{M} 
\newcommand{\stiffFEM}{K} 

\newcommand{\Atmo}{\texttt{Atmo}}

\begin{document}

   \title{Learned infinite elements for helioseismology}

   \subtitle{Learning transparent boundary conditions for the solar atmosphere}

   \author{D. Fournier
          \inst{1}
          \and
          T. Hohage\inst{2,1}
	  \and
	  J. Preuss\inst{3}
   \and
   L. Gizon\inst{1,4}
          }

   \institute{Max-Planck-Institut f\"ur Sonnensystemforschung,
              Justus-von-Liebig-Weg 3, 37077 G\"ottingen, Germany \\
              \email{fournier@mps.mpg.de}
         \and
             University of G\"ottingen, Institute for numerical and applied mathematics,
             Lotzestr. 16-18, 37083 G\"ottingen, Germany
	 \and
             University College London, Department of Mathematics,
             Gower Street, WC1E 6BT, London, United Kingdom \\
             \email{j.preuss@ucl.ac.uk}
    \and
             University of G\"ottingen, Institute for Astrophysics and Geophysics, Friedrich-Hund-Platz 1, 37077 G\"ottingen, Germany
             }

   \date{Received \today; accepted xx xx, xxxx}


  \abstract
   {
   Acoustic waves in the Sun are affected by the atmospheric layers, but this region is often ignored in forward models due to the increase in computational cost.
   }
   {
   The purpose of this work is to take into account the solar atmosphere without increasing significantly the computational cost.
   }
   {
   We solve a scalar wave equation that describes the propagation of acoustic modes inside the Sun using a finite element method. The boundary conditions used to truncate the computational domain are learned from the Dirichlet-to-Neumann operator, that is the relation between the solution and its normal derivative at the computational boundary.  These boundary conditions may be applied at any height above which the background medium is assumed to be radially symmetric.
   }
   {
   Taking into account the atmosphere is important even for wave frequencies below the acoustic cut-off. In particular, the mode frequencies computed for an isothermal atmosphere differ by up 10~$\mu$Hz from those computed for the VAL-C atmospheric model.
   We show that learned infinite elements lead to a numerical accuracy similar to that obtained for a traditional radiation boundary condition. Its main advantage is to reproduce the solution for  any radially symmetric atmosphere to a very good accuracy at a low computational cost.
   }
   {
   This work emphasizes the importance of including atmospheric layers in helioseismology and proposes a computationally efficient method to do so.
   }


   \maketitle
%

\section{Introduction}

Accurate forward modelling of wave propagation is an essential step in the interpretation of solar oscillations. Modelling acoustic waves requires appropriate boundary conditions. The surface boundary conditions are especially important at high-frequencies, for waves that are not trapped in the interior.

The most common boundary condition used in helioseismology  is the free-surface boundary condition, such that the Lagrangian perturbation of the pressure vanishes at the surface. Such assumption \citet{LO67} ensures the self-adjointness of the adiabatic equations of stellar oscillations. It is commonly used in helioseismology but does not allow for the treatment of high-frequency waves. An  improvement is to match the solar model with an isothermal atmosphere, which allows waves above the acoustic cut-off frequency to propagate into the atmosphere \citep[see e.g.][]{Unno89}. This type of  atmosphere is often used to compute the eigenvalues of solar oscillations \citep{C08,TT13} and has been extensively studied \citep{FL17,BC18,BFP19} for the simplified scalar problem
\begin{equation}
    - \frac{\sigma^2}{\rho c^2} \psi - \nabla \cdot \left( \frac{1}{\rho} \nabla \psi \right)
    = s, \label{eq:scalar_wave}
\end{equation}
where $\psi = \rho c^2 \nabla \cdot \boldsymbol{\xi}$ is proportional to the divergence of the wave displacement $\boldsymbol{\xi}$. In this equation, $\rho$ is the density, $c$ the sound speed, and $\sigma^2 = \omega^2 + 2 \ii \omega \gamma$ is the complex frequency. 
Such an equation is obtained from the equations of stellar oscillations \citep{LO67} after neglecting gravity and background flows. \citet{BC18} derived boundary conditions that are relatively easy to implement and often offer sufficient accuracy for applications in time-distance helioseismology \citep{FL17}.

\begin{figure}
\begin{tikzpicture}
  \path[fill=orange!30,even odd rule]
    (0,0)circle[radius=4.]
    (0,0)circle[radius=3]
  ;

\draw[color=red](0,0) circle(4);
\draw[color=red, dashed](0,0) circle(3);
\draw[black,->](0,-4.5) -- (0,4.5);
\node[color=black] at (0.,4.7) {$\hat{z}$};
\node[color=red] at (0.7,3.2) {$r_a$};
\node[color=red] at (0.7,4.2) {$r_t$};
\node[color=red] at (2.5,2.) {$\Gamma$};

\node[color=black] at (3.5,0.5) {$\rho(r)$};
\node[color=black] at (3.5,0.) {$c(r)$};
\node[color=black] at (3.45,-0.5) {$\gamma(r,\omega)$};
\node[color=black] at (-1.5,1.) {$\rho(\mathbf{r})$,};
\node[color=black] at (-1.5,0.5) {$c(\mathbf{r})$,};
\node[color=black] at (-1.5,0.) {$\gamma(\mathbf{r},\omega)$,};
\node[color=black] at (-1.5,-0.5) {$s(\mathbf{r}).$};
\draw[blue,dashed](0,0) -- node[above,rotate=30] {r} (1,1.5);
\draw[blue] (0,0) ++(90:0.5) arc (90:60:0.5);
\node[color=blue] at (0.15,0.65) {$\theta$};
\node[color=blue,rotate=45] at (1,1.5) {+};
\end{tikzpicture}
\caption{In the solar interior (up to $r_a$), the background medium is characterized by its density, sound speed, attenuation, and sources that can depend on the 3D vector $\mathbf{r}$. In the atmosphere ($r_a \leq r \leq r_t$), the background medium is radially symmetric and does not contain sources. The boundary condition is usually applied at $r_t$ but can be directly applied at $r_a$ using the learned infinite elements.}
\label{fig:Sun-sketch}%
\end{figure}
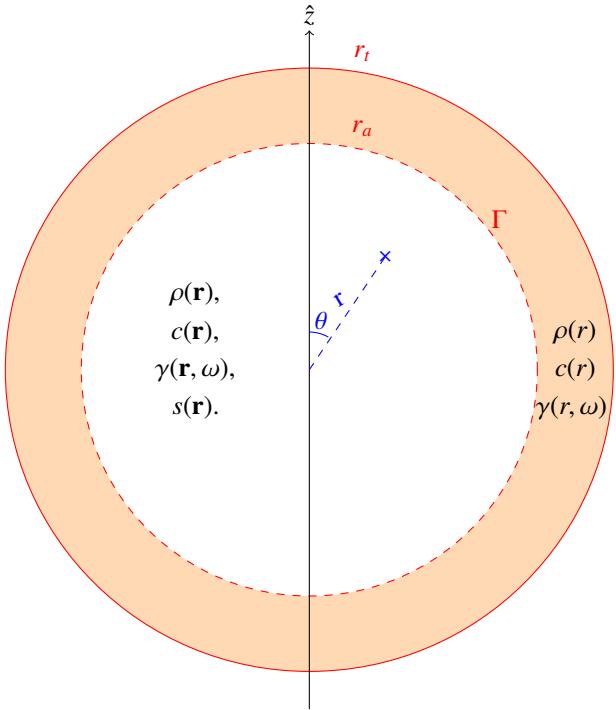

However, all these boundary conditions are tied to a particular model of atmosphere which needs to be simple enough so that boundary conditions can be derived analytically. In this paper, we overcome this restriction by using learned infinite elements \citep{HLP21}  which emulates a boundary condition for any type of radially symmetric profile in the atmosphere. It allows to truncate the computational domain at any height (above which we are not interested in the solution, e.g., the measurement height) even if the background parameters are still varying. Thus, any 1D model of atmosphere can be used on top of a background solar model without significantly increasing the computational cost. As high-frequency waves propagate into the atmosphere and can be reflected back in the interior due to the high increase of temperature in the chromosphere, a proper modeling of the atmosphere is important for helioseismology. The influence of the atmosphere on the time-distance diagram for high-frequency waves has been reported for example by \citet{JO97,FL17}. However, we will see that the atmosphere also influences waves with frequencies well below the acoustic cut-off.

The traditional approaches to derive the boundary conditions for Eq.~\eqref{eq:scalar_wave} are presented in Sect.~\ref{section:forward} while the new approach using learned infinite elements is explained in Sect.~\ref{section:learned}. The advantages of this approach are numerically illustrated in Sect.~\ref{section:numerical} while some possible extensions of this work are sketched in Sect.~\ref{section:conclusion}.


\section{Scalar wave equation and transparent boundary conditions} \label{section:forward}

We would like to solve Eq.~\eqref{eq:scalar_wave} in the geometry sketched in Fig.~\ref{fig:Sun-sketch}. In the interior ($r < r_a$), all the coefficients of Eq.~\eqref{eq:scalar_wave} can depend on the 3D vector $\mathbf{r}$. On the other hand, the atmosphere ($r_a \leq r \leq r_t$) is assumed to be source-free and spherically symmetric (no background flow). This greatly simplifies the solution of
Eq.~\eqref{eq:scalar_wave} in this region. Indeed, let us introduce the ansatz
 \begin{equation}\label{eq:decomp-int-ext}
 \psi =
      \begin{cases}
         \psi^{\mathrm{i}} \quad  \text{ for } r \leq r_a , \\
         \psi^{\mathrm{e}} \quad \text{ for } r_a \leq r \leq r_t,
 \end{cases}
 \end{equation}
 and decompose the solution in the exterior into spherical harmonics
 \begin{align}
    \psi^{\mathrm{e}}(r,\theta,\phi) &= \sum_{\ell,m} \psi_{\ell m}^{\mathrm{i}}(r_a) \psi_{\ell m}^{\mathrm{e}}(r) Y_\ell^m(\theta,\phi). \label{eq:scalar_wave_modal}
\end{align}
Here, $\psi_{\ell m}^{\mathrm{i}}(r_a) $ are the coefficients in the spherical harmonics expansion of $\psi^{\mathrm{i}} $ at $r=r_a$, i.e.\
\begin{equation}
           \psi^{\mathrm{i}}|_{r=r_a} = \sum_{\ell,m} \psi_{\ell m}^{\mathrm{i}}(r_a) Y_\ell^m(\theta,\phi).
           \label{eq:scalar_wave_modal_inner}
\end{equation}
The continuity of $\psi$ decomposed as in Eq.~\eqref{eq:decomp-int-ext} is ensured at $r = r_a$ if we require
\begin{equation}\label{eq:ODE-continuity-scalar}
    \psi_{\ell m}^{\mathrm{e}}(r_a) = 1.
\end{equation}
By inserting ansatz~\eqref{eq:scalar_wave_modal} for $ \psi^{\mathrm{e}} $ into Eq.~\eqref{eq:scalar_wave}, we obtain thanks to the radial-symmetry of $\rho$ and $c$ the ordinary differential equation (ODE)
\begin{equation}\label{eq:dtn-ODE}
    - \left( \frac{\sigma^2}{c^2} - k_h^2 \right) \psi_{\ell m}^{\mathrm{e}} - \frac{\rho}{r^2} \frac{\dd}{\dd r} \left( \frac{r^2}{\rho} \frac{\dd \psi_{\ell m}^{\mathrm{e}}}{\dd r} \right) = 0, \quad  r_a < r < r_t.
\end{equation}
Here, we have used that the source $s$ vanishes in the atmosphere and introduced the horizontal wavenumber $k_h := \sqrt{\ell(\ell+1)}/r$.
Deriving a similar equation for $\psi^{\mathrm{i}} $ in the interior is only possible, if the assumption of spherical symmetry is extended to the entire Sun, which is usually not possible due to the presence of heterogeneity.

Before going into further details on how Eq.~\eqref{eq:dtn-ODE} is solved, let us discuss the problem of matching the interior solution $ \psi^{\mathrm{i}} $ and exterior solution $ \psi^{\mathrm{e}} $ in a meaningful way. To this end, we will require that
the normal derivatives $\partial_n$ (radial derivative in the case of a spherical star) match at the interface, that is
\begin{equation}\label{eq:matching-cond}
-\partial_n \psi^{\mathrm{i}} =  -\partial_n \psi^{\mathrm{e}} \; \text{at } r = r_a.
\end{equation}
In view of Eq.~\eqref{eq:scalar_wave_modal}, this means that
\begin{equation}\label{eq:exact-Dtn-coupling-exterior}
    -(\partial_n \psi^{\mathrm{i}})|_{r=r_a} = \sum_{\ell,m} \dtn_\ell \psi_{\ell m}^{\mathrm{i}}(r_a) Y_\ell^m =: \DtN \psi^{\mathrm{i}}|_{r=r_a},
\end{equation}
where $\DtN$ is called the Dirichlet-to-Neumann operator and
\begin{equation}\label{eq:dtn-nr-exact}
\dtn_\ell := -\frac{\dd \psi_{\ell m}^{\mathrm{e}}}{\dd r}(r_a)
\end{equation}
are its coefficients in the basis of spherical harmonics.
Note that Eq.~\eqref{eq:dtn-ODE} does not depend on $m$, which implies that $\dtn$ is a function of $\ell$ only.

To summarize, in order to solve Eq.~\eqref{eq:scalar_wave} in the geometry of Fig.~\ref{fig:Sun-sketch} we proceed as follows:
\begin{itemize}
    \item Solve Eq.~\eqref{eq:dtn-ODE} with boundary condition given in Eq.~\eqref{eq:ODE-continuity-scalar} to determine $\DtN$, i.e.\ the complex numbers $\dtn_\ell $. This task is discussed in Sect.~\ref{ssection:atmo-ODE} in more detail.
    \item Then solve Eq.~\eqref{eq:scalar_wave} for $\psi = \psi^{\mathrm{i}}$ with boundary condition given by Eq.~\eqref{eq:exact-Dtn-coupling-exterior}. We discuss this task in Sect.~\ref{ssection:DtN-impl}.
\end{itemize}

\subsection{Atmospheric model}\label{ssection:atmo-ODE}
To solve Eq.~\eqref{eq:dtn-ODE}, we first have to discuss the atmospheric model. A commonly used model is an isothermal atmosphere where the sound speed is  constant and the density decreases exponentially. We will refer to this model as \Atmo \ \citep{FL17}.
However, this model does not take into account the high increase of temperature in the transition region resulting in an increase of sound speed by one order of magnitude. To this end, we use the empirical VAL-C model of atmosphere \citep{VAL81}. For the interior, we use the standard solar model~S \cite{CD96}. A smooth transition between the two models is used between 400 and 600~km above the surface \citep[see Fig. 2 in][]{FL17}. We choose this model to validate our approach, but the method presented in this paper is general and can be applied to any radially-symmetric atmospheric model.

Let us first discuss the \texttt{Atmo} model in more detail. In this case the atmosphere is not bounded, that is $r_t = \infty $. However, the computational domain needs to be truncated to solve this problem numerically. To derive outgoing solutions, it is useful to recast Eq.~\eqref{eq:dtn-ODE} into a Schr\"odinger-type equation using the change of variable $\phi_{lm}^e = r \rho^{-1/2} \psi_{lm}^e$. Equation~\eqref{eq:dtn-ODE} becomes
\begin{equation}
    \frac{\dd^2 \phi_{lm}^e}{\dd r^2} + \left( \frac{\sigma^2 - \omega_c^2}{c^2} - k_h^2 \right) \phi_{lm}^e = 0,
\end{equation}
where $\omega_c$ is the cut-off frequency defined as
\begin{equation}
    \frac{\omega_c^2}{c^2} = \frac{1}{4H^2} \left(1 - 2 \frac{\dd H}{\dd r} + \frac{4 H}{r} \right),
\end{equation}
and $H$ is the density scale height
\begin{equation}
    H = - \left( \frac{\dd \ln \rho}{\dd r} \right)^{-1}.
\end{equation}
The cut-off frequency is an important quantity as waves with frequencies above $\omega_c$ are propagating into the atmosphere while the ones below are mostly reflected back into the Sun. The study of waves above the cut-off frequency requires an appropriate treatment of the boundary conditions but we will see that lower frequencies are also impacted by the choice of atmosphere.

In the special case of the \Atmo \ model, an analytic expression of the $\dtn$ was obtained by \citet{BFP19b} in terms of the Whittaker functions $W$
\begin{equation}\label{eq:dtn_atmo}
\dtn^{\texttt{Atmo}}_\ell =   \left( \frac{1}{r_a} + \frac{1}{2H} \right) +  2 i \kk \frac{ W^{\prime}_{-i / (2 \kk H), \ell +  \frac{1}{2} }(  -2i \kk r_a )   }{ W_{- i / (2 \kk H) , \ell +  \frac{1}{2}  } (  -2i \kk r_a) },
\end{equation}
where the wavenumber $\kk$ is
\begin{equation}
    \kk^2 = \frac{\sigma^2}{c^2} - \frac{1}{4 H^2}.
\end{equation}
The Whittaker functions are defined as the outgoing solutions of
\begin{equation}
    \frac{\dd^2 W_{\kappa,\mu}}{\dd r^2} + \left( - \frac{1}{4} + \frac{\kappa}{r} + \frac{1/4 - \mu^2}{r^2} \right) W_{\kappa,\mu} = 0.
\end{equation}
and can be evaluated with libraries such as \texttt{arb} \citep{FJ17}.

Giving an explicit formula for $ \dtn^{\texttt{VAL-C}}_\ell $ corresponding to the VAL-C model is not possible. In this case we have to numerically solve  Eq.~\eqref{eq:dtn-ODE} on the interval $[r_a,r_t]$. As the data for the VAL-C model stops at about $2.5$Mm above the surface, the right end $r_t$ of the interval is finite. We impose a homogeneous Dirichlet boundary condition at $r_t$ to complement the boundary condition \eqref{eq:ODE-continuity-scalar} at $r_a$. Due to the high increase of temperature in the transition region, waves are not reaching the end of the domain, and the choice of boundary condition applied after the end of the VAL-C model does not cause significant differences in the obtained solutions. Equation~\eqref{eq:dtn-ODE} is a two-point boundary value problem which can be solved in various ways, see, e.g., \cite{K92book}. We prefer to use the finite element method (FEM) since it is already employed for discretizing the solar interior and allows for a convenient evaluation of the derivative of $ \psi_{\ell m}$ at $r_a$ required to obtain $ \dtn^{\texttt{VAL-C}}_\ell $ via  Eq.~\eqref{eq:dtn-nr-exact}.


\subsection{Implementing the $\DtN$ condition}\label{ssection:DtN-impl}

Let us now discuss how to implement the boundary condition given in Eq.~\eqref{eq:exact-Dtn-coupling-exterior} for Eq.~\eqref{eq:scalar_wave}. We briefly discuss the special case of a spherically symmetric background before moving on to the general setting.

\subsubsection{Spherically symmetric background}\label{ssection:spherically-symmetric-background}

In a spherically symmetric background, we can also decompose the interior
solution $\psi^{\mathrm{i}} $ into spherical harmonics analogously to Eq.~\eqref{eq:scalar_wave_modal} and derive the following ODE for its coefficients:
\begin{equation}
    - \left( \frac{\sigma^2}{c^2} - k_h^2 \right) \psi_{\ell m}^{\mathrm{i}} - \frac{\rho}{r^2} \frac{\dd}{\dd r} \left( \frac{r^2}{\rho} \frac{\dd \psi_{\ell m}^{\mathrm{i}}}{\dd r} \right) = \rho s_{lm}, \quad  0 \leq r \leq r_a, \label{eq:scalar_wave1d}
\end{equation}
where $s_{\ell m}$ are the coefficients of the source term $s$ in spherical harmonics.
This allows for an exact implementation of the boundary condition in Eq.~\eqref{eq:exact-Dtn-coupling-exterior} in the form
\begin{equation}\label{eq:dtn-modal-impl}
  -\frac{\dd \psi_{\ell m}^{\mathrm{i}}}{\dd r}(r_a) =  \dtn_\ell \psi_{\ell m}^{\mathrm{i}}(r_a).
\end{equation}

\subsubsection{General background}\label{sssection:genral-background}

If the background is not spherically symmetric, for example due to the presence of background flow, then an exact implementation of Eq.~\eqref{eq:exact-Dtn-coupling-exterior} is no longer feasible. We have to replace $\DtN$ in Eq.~\eqref{eq:exact-Dtn-coupling-exterior}
by a simpler expression which approximates its action on the spherical harmonics, i.e.\ $ \DtN Y_\ell^m = \dtn_\ell  Y_\ell^m  $,  as good as possible.
To this end, a fairly popular choice is
\begin{equation}\label{eq:dtn-affine-linear}
    \DtN^{0} \psi^{\mathrm{i}} = \alpha \psi^{\mathrm{i}} - \beta \Delta_{\Gamma} \psi^{\mathrm{i}},
\end{equation}
for some $\alpha,\beta \in \mathbb{C}$, where $-\Delta_{\Gamma}$ is the horizontal Laplacian on the boundary $\Gamma := \{\mathbf{r} , ||\mathbf{r}||=r_a\}$ fulfilling
\begin{equation}\label{eq:spherical-harmonics-eigenfunctions}
-\Delta_{\Gamma}  Y_\ell^m = \lambda_{\ell}Y_\ell^m
\end{equation}
 for $ \lambda_{\ell} = \ell (\ell+1) / r_a^2$. This expression is easy to implement using the FEM. Upon multiplying Eq.~\eqref{eq:scalar_wave} by a test function $\phi^{\mathrm{i}}$ and integrating by parts, the normal derivative $-\partial_n \psi^{\mathrm{i}}$ on $\Gamma$ appears naturally:
 \begin{equation}\label{eq:scalar_wave_weak}
 \int\limits_{ \{ r < r_a \} } \left( \frac{1}{\rho} \nabla  \psi^{\mathrm{i}} \nabla  \phi^{\mathrm{i}} - \frac{\sigma^2}{\rho c^2} \psi^{\mathrm{i}} \phi^{\mathrm{i}} \right)  \; \mathrm{d}V - \int\limits_{\Gamma} \frac{1}{\rho} \partial_n \psi^{\mathrm{i}} \phi^{\mathrm{i}} \; \mathrm{d}S = \int s \phi^i dV.
 \end{equation}
 It can then be replaced using the approximation $-\partial_n \psi^{\mathrm{i}} \approx  \DtN^{0} \psi^{\mathrm{i}} $.
This amounts to replacing the second integral in Eq.~\eqref{eq:scalar_wave_weak} by
\begin{equation}\label{eq:impl-low-order-FEM}
\int\limits_{\Gamma} \frac{1}{\rho} \left(  \alpha \psi \phi + \beta \nabla_{\Gamma} \psi \nabla_{\Gamma} \phi  \right) \; \mathrm{d}S.
\end{equation}
Note that $-\Delta_{\Gamma}$ has been integrated by parts leading to the surface gradients $\nabla_{\Gamma} = (I - n \cdot n^T) \nabla $.

The constants $\alpha$ and $\beta$ should be determined  such that the affine function
$\lambda_{\ell} \mapsto \dtn^{0}_{\ell} $ in
\begin{equation}\label{eq:DtN-lin}
    \DtN^{0} Y_\ell^m = \dtn^{0}_{\ell} Y_\ell^m, \quad \dtn^{0}_{\ell} = \alpha + \beta \lambda_{\ell}
\end{equation}
fits the exact function $\dtn_{\ell}$ in Eq.~\eqref{eq:dtn-nr-exact} optimally.
\begin{figure}
\centering
\includegraphics[width=.5\textwidth]{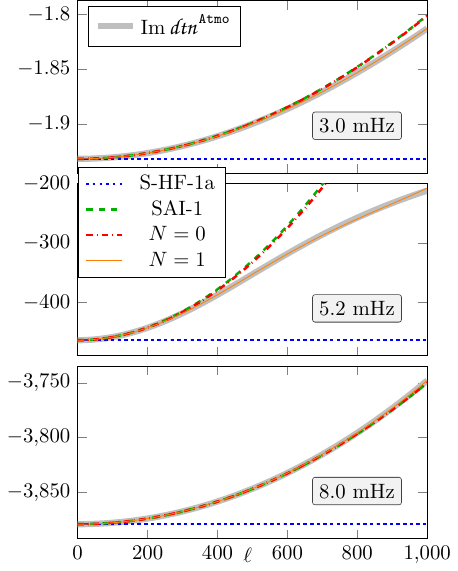}
	\caption{Comparison of $\dtn$ approximations provided by different transparent boundary conditions at $3.0, 5.2$ and $8.0$~mHz for the \Atmo \ model at $r_t = 510$~km. The reference is given by Eq.~\eqref{eq:dtn_atmo}, and the boundary conditions S-HF-1a and SAI-1 from \citet{BFP19} are recalled in Eqs.~\eqref{eq:S-HF-1a}~and~\eqref{eq:SAI-1}. The learned infinite elements boundary conditions for $N=0$ and $N=1$ are obtained by solving the minimisation problem defined in Eq.~\eqref{eq:NLLSE}.   }
\label{fig:dtn-approx-low-order}%
\end{figure}
Two choices for the \texttt{Atmo} model derived in \cite{BFP19} are
\begin{align}
    \dtn^{\rm S-HF-1a}  &= \left[  \left( \frac{1}{r_a} - \frac{1}{2H} \right)  - \ii \kk + \frac{\ii}{\kk H r_a } \right], \label{eq:S-HF-1a} \\
    \dtn^{\rm SAI-1} \psi &= \left[  \left( \frac{1}{r_a} + \frac{1}{2H} \right) - \ii \kk \left( 1 - \frac{1}{\kk^2 H r_a } \right)^{1/2} \right]  \nonumber \\
    &- \frac{\ii}{2 \kk}  \left( 1 - \frac{1}{\kk^2 H r_a} \right)^{-1/2} \lambda_{\ell}. \label{eq:SAI-1}
\end{align}
A representation of the exact $\dtn$ and these two approximations is given in Fig.~\ref{fig:dtn-approx-low-order} for three typical frequencies: below, around and above the cut-off frequency. The S-HF-1a boundary condition is of order 0 and does not depend on the harmonic degree ($\beta =0$) while SAI-1 is an affine approximation of the $\dtn$ and scales as $\ell(\ell+1)$. For low harmonic degrees $\ell < 100$, both boundary conditions yield good approximations to  the $\dtn$ but for larger values of $\ell$ an affine approximation is necessary. While for frequencies below and above the cut-off frequency the affine model is a very good approximation of the exact $\dtn$, this is not the case around the cut-off frequency for large values of $\ell$. However, the affine approximation is very good for $\ell \leq 300$ which is the range often considered in helioseismology to study large-scale flows \citep[see e.g.][for meridional flow]{GC20}.

Finally, we would like to point out that the derivation of the conditions given in Eqs.~\eqref{eq:S-HF-1a}~and~\eqref{eq:SAI-1} is based on the assumption that sound speed and density scale height are constant in the atmosphere, which is only valid in the isothermal case. It is not known how to derive analytical expressions for the coefficients $\alpha$ and $\beta$ in ~\eqref{eq:dtn-affine-linear} for other atmospheric models like VAL-C.
Thus, for such models one would need to resort to meshing the entire atmosphere, i.e.\  place the boundary $\Gamma$ at $r=r_t$ instead of $r=r_a$.
This results in a dramatic increase of computational cost and memory requirements as the wavelength is small in the atmosphere due to the low sound speed. Note that an infinite atmosphere ($r_t = \infty$) could not even be treated using the finite element method which requires the computational domain to be finite.


\section{Learned infinite elements}\label{section:learned}

The method of learned infinite elements was introduced by \citet{HLP21} to allow for an accurate and efficient approximation of the exact $\DtN$ boundary condition given in Eq.~\eqref{eq:exact-Dtn-coupling-exterior}. It generalizes the approach introduced in Sect.~\ref{sssection:genral-background} in two aspects. Firstly, it allows to obtain optimal parameters $\alpha$ and $\beta$ in Eq.~\eqref{eq:DtN-lin} for any radially symmetric model of the atmosphere as will be described in Sect.~\ref{ssection:lie-low-order}. Secondly, it permits to efficiently implement rational approximations of $\dtn$, i.e.\ we can replace Eq.~\eqref{eq:DtN-lin} by
\begin{equation}\label{eq:DtN-N}
    \DtN^{N} Y_\ell^m = \dtn^{N}_{\ell} Y_\ell^m,
\end{equation}
where $\dtn^{N}_{\ell} $ is a rational function of $\lambda_{\ell}$ with $N$ simple poles. Thanks to the superior approximation qualities of rational functions, this allows to obtain a very accurate fit of $\dtn$ even when its behavior as a function of $\lambda_{\ell}$ is fairly complicated. We explain this advantage in detail in Sect.~\ref{section:high-order-learnedIE}.

\subsection{Affine  approximation}\label{ssection:lie-low-order}

We recall that the objective is to find parameters $\alpha$ and $\beta$ which minimize the deviation $\vert \dtn^{0}_{\ell} - \dtn_{\ell} \vert = \vert \alpha + \beta \lambda_{\ell} - \dtn_{\ell} \vert$ across a range of $0 \leq \ell  \leq L$ for some $L \geq 1$.
To this end, it seems natural to set up a least-squares fit. Moreover, in helioseismology it is common practice to apply filters which localize in $\ell$ over a certain range of primary interest. We can take this into account here by multiplying the deviation additionally with some non-negative weights $w_{\ell}$. Hence, we want to find $\alpha,\beta \in \mathbb{C}$ minimizing
\begin{equation}\label{eq:LLSE}
\left\Vert \Lambda
\begin{pmatrix}
 \alpha \\
\beta
\end{pmatrix}
- b \right\Vert_2^2,
\; \text{ for }
\Lambda = \begin{pmatrix}
 w_0 & w_0 \lambda_0 \\
\vdots & \vdots \\
 w_L & w_L \lambda_L
\end{pmatrix},
\;
b = \begin{pmatrix}
 w_0 \dtn_{0} \\
\vdots  \\
 w_L \dtn_{L}
\end{pmatrix},
\end{equation}
where $\Vert \cdot \Vert_2 $ denotes the Euclidean norm.
The minimal norm solution is given by
\begin{equation}\label{eq:LLSE-sol}
 \begin{pmatrix}
 \alpha \\
\beta
\end{pmatrix} = \Lambda^{\dagger} b,
\end{equation}
where $\Lambda^{\dagger}:=(\Lambda^*\Lambda)^{-1}\Lambda^*$ denotes the pseudoinverse of $\Lambda$. The explicit expression of the coefficients $\alpha$ and $\beta$ as a function of $w_\ell$ and $\dtn_\ell$ is given in Appendix~\ref{section:minimization}.

\subsection{Rational approximation}\label{section:high-order-learnedIE}

It is straightforward to generalize the least-squares approach introduced in the previous section to the rational approximation of $\dtn$ announced in Eq.~\eqref{eq:DtN-N}. A rational function of $\lambda_{\ell}$ with $N$ simple poles is given by
 \begin{equation}\label{eq:dtn-N}
 \dtn^N_{\ell}  = \dstiffr_{00} + \lambda \dmassr_{00} - \sum\limits_{j=1}^{N}{ \frac{(\dstiffr_{0j} + \lambda \dmassr_{0j})^2 }{  \dstiffr_{jj} + \lambda  }     }
 \end{equation}
 provided $\dstiffr_{0j} \neq \dmassr_{0j} \dstiffr_{jj} $ at $j= 1, \ldots,N$ holds. This ansatz is justified by the result established in \citep[Proposition 3.13]{JP_PHD_2021} which ensures that there exist complex numbers $\dstiffr_{ij}$ and $\dmassr_{ij}$ such that the deviation $ \vert  \dtn^N_{\ell} - \dtn_{\ell} \vert$ for $\ell \leq L$ decreases exponentially fast as $N$ increases.
 In analogy with Sect.~\ref{ssection:lie-low-order} we obtain  $\dstiffr_{ij}$ and $\dmassr_{ij}$ in practice by minimization of the least-squares functional
 \begin{equation}
   \frac{1}{2} \sum\limits_{\ell =0}^{L}  w_{\ell}^2 \vert  \dtn^N_{\ell} - \dtn_{\ell} \vert^2.
   \label{eq:NLLSE}
 \end{equation}
 We recognize that problem~\eqref{eq:NLLSE} reduces to problem~\eqref{eq:LLSE}
 for $N=0$ and the solution $(\dstiffr_{00}, \dmassr_{00}) = (\alpha,\beta) $ is given by Eq.~\eqref{eq:LLSE-sol}. For $N >0$ on the other hand, problem~\eqref{eq:NLLSE} becomes non-linear and no longer allows for an explicit solution. The Levenberg–Marquardt algorithm  \citep[see e.g.][]{NW06} is a popular option to solve such problems iteratively. In practice, it is observed to perform well for minimizing~\eqref{eq:NLLSE}. For example, Fig.~\ref{fig:dtn-approx-low-order} compares the affine ($N=0$) and first-order rational approximation ($N=1$) to the established approximations for the \texttt{Atmo} model.
 The affine approximation ($N=0$) obtained by solving the least-squares problem can at least reproduce the accuracy of the SAI-1 condition. To achieve a better fit of $\dtn$, the rational approximation ($N=1$) can be used. The improvement in accuracy is clearly visible at $3.0$  and $5.2$~mHz.

\begin{figure}
\centering
\includegraphics[width=.5\textwidth]{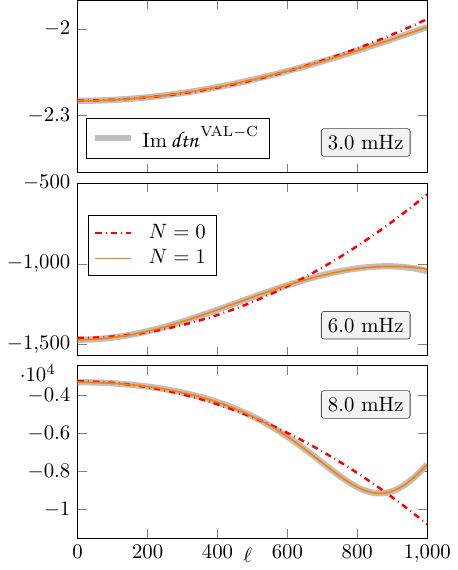}
\caption{
Exact $\dtn$ for the VAL-C atmospheric model and its approximations using learned infinite elements for $N=0$ and 1. The boundary condition is applied directly at the solar surface $r_a = R_\odot$. }
\label{fig:dtn-VALC}%
\end{figure}

The big advantage of learned IE is to be applicable to any spherically-symmetric exterior model (by changing the data vector $b$ in ~\eqref{eq:LLSE}), whereas the radiation boundary conditions from \citet{BC18} are tied to the \texttt{Atmo} model. A representation of the $\dtn$ operator for the VAL-C model applied directly at the solar surface ($r_a = R_\odot$) is shown in Fig.~\ref{fig:dtn-VALC}. Its behaviour is more complicated than for the isothermal atmosphere (Fig.~\ref{fig:dtn-approx-low-order}) and cannot be approximated by a low-order boundary condition. The learned infinite elements with $N=0$ only gives a crude approximation of the $\dtn$ and at least $N=1$ is necessary to represent the variations with $\ell$.

 \subsection{Infinite elements}\label{section:infel}
 We call the method based on the rational $\DtN^{N}  $ approximation in Eq.~\eqref{eq:DtN-N} with $\dtn^N$ obtained from solving the minimization problem in Eq.~\eqref{eq:NLLSE} `learned infinite elements'. The term `learned' describes the process of acquiring the coefficients $A_{ij}$ and $B_{ij}$ of the rational approximation from the given data describing the atmosphere, i.e. the numbers $\dtn_{\ell}$ determined by Eq.~\eqref{eq:dtn-ODE}.
 \begin{figure}
\centering
\includegraphics[scale=0.65]{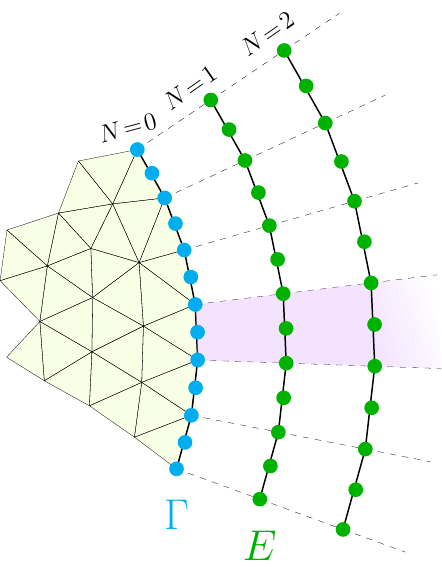}
	\caption{Schematic illustration of learned infinite elements. The domain is meshed in the solar interior (up to $\Gamma$) and degrees of freedom are defined at the boundary $\Gamma$ as for classical finite elements. For $N>0$, additional degrees of freedom are located in the exterior (see more details in Appendix~\ref{section:FEM-impl}).    }
\label{fig:learnedIE-sketch-Sun}%
\end{figure}
 The term `infinite elements' refers to the algebraic structure which arises when implementing  Eq.~\eqref{eq:DtN-N} in the context of the finite element method. The main idea is illustrated in Fig.~\ref{fig:learnedIE-sketch-Sun}. For $N=0$ we add the surface integral in  Eq.~\eqref{eq:impl-low-order-FEM} involving the degrees of freedom (\dofs) on $\Gamma$ weighted by the coefficients $\dstiffr_{00}$ and $\dmassr_{00}$. For $N=1$ we duplicate the \dofs \ of $\Gamma$ and add them in the form of an additional layer on top of the existing discretization. The corresponding matrix entries are determined from surface integrals like in Eq.~\eqref{eq:impl-low-order-FEM} weighted with the coefficients $\dstiffr_{ij}$ and $\dmassr_{ij}$.  This approach is computationally efficient as it preserves the sparsity structure of the finite elements and only very few additional such layers (typically $N<5$, and often $N=1$) are required to achieve high accuracy. We refer the reader to Appendix~\ref{section:FEM-impl} for a detailed description of the implementation in the context of FEM and to Sect.~\ref{ssection:comp-cost} for a discussion of computational costs.

\begin{figure*}
\centering
\includegraphics[width=\textwidth]{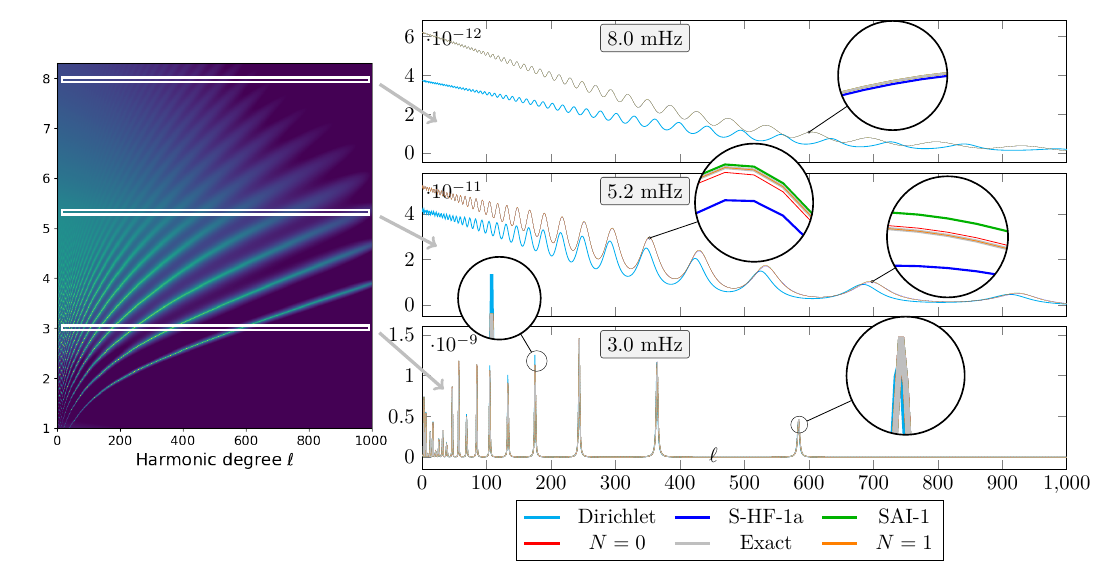}
\caption{Left: Power spectrum for the \texttt{Atmo} model using the exact $\dtn$. Right: Cuts at different frequencies showing the power spectra obtained with different transparent boundary conditions.}
\label{fig:power-Atmo-freq-cuts}%
\end{figure*}

\section{Numerical experiments} \label{section:numerical}

In order to test the accuracy of the proposed boundary conditions in terms of physical quantities relevant in helioseismology, we compare power spectrum, cross-covariance, and travel times. We explain first how the Green's function is computed before showing its relation to the power spectrum using the setup from \citet{GB17}.

\subsection{Green's function}

The reference Green's function is obtained
as described in Sect.~\ref{ssection:spherically-symmetric-background}, where the source is a delta function $\delta(r-r_0)$ located at the surface $r_0 = R_{\odot}$. The attenuation model is based on the observed linewidths and follows the power
law from \citet{GB17}. To obtain the exact boundary condition, we use $\dtn_\ell = \dtn^{\texttt{Atmo}}_\ell$  (Eq.~\eqref{eq:dtn-modal-impl}) for the \Atmo \ model and $\dtn_\ell = \dtn^{\texttt{VAL-C}}_\ell$ for the VAL-C model.  The different transparent boundary conditions are realized by replacing the exact $\dtn_\ell$ numbers in Eq.~\eqref{eq:dtn-modal-impl} by their respective approximations, i.e.\ $\dtn^{\rm S-HF-1a}_{\ell},\dtn^{\rm SAI-1}_{\ell}$ or $\dtn^N_{\ell}$. For the approximations of the \Atmo \ model, we apply the boundary condition at $r_a = R_\odot + 510$~km as the radiation boundary conditions were derived under the hypothesis of constant sound speed and density scale height. On the other hand, learned IE can be applied at any height above which the medium is radially symmetric and without sources. Thus, we apply the boundary condition directly at the solar surface ($r_a = R_\odot$) for the VAL-C atmosphere leading to a drastic reduction of the computational costs (see Sect.~\ref{ssection:comp-cost}).

\subsection{Power spectrum}

 If the sources in the Sun are equipartitioned, \citet{GB17} showed that the power spectrum is related to the imaginary part of the Green's function
\begin{equation}\label{eq:power_spectrum_modal}
	\mathcal{P}_{\ell}(\omega) = \frac{\Pi(\omega)}{2 \omega} \textrm{Im} \left[ G_{\ell}(r_0, r_0,\omega) \right] \mathcal{F}_\ell,
\end{equation}
where $r_0$ is the observation height. The function $\Pi(\omega)$ controls the frequency distribution of the sources and is chosen as
\begin{equation}\label{eq:source_power}
	\Pi(\omega) = \left( 1 + \left( \frac{  \omega  - \omega_{0} }{ \sigma_{0} }  \right)^2  \right)^{-1} \: \text{with }
	\frac{\omega_{0}}{2 \pi} = 3.3 \text{ mHz },
        \frac{\sigma_{0}}{2 \pi} = 0.6 \text{ mHz }
\end{equation}
so that the distribution of the power spectrum with frequency ($\sum_\ell \mathcal{P}_\ell(\omega)$) matches reasonably well the observations. Similarly, $\mathcal{F}_\ell$ is chosen so that the distribution of the power spectrum with the harmonic degree ($\sum_\omega \mathcal{P}_\ell(\omega)$) agrees with the HMI power spectrum. We fit a polynomial of order 4
\begin{equation}\label{eq:Power-fit}
    \mathcal{F}_\ell = \sum_{i=0}^4 f_i \bigl(\ell / L_{\rm max} \bigr)^i,
\end{equation}
where $L_{\rm max} = 1000$ is the maximum harmonic degree used in this study and the coefficients $f_i$ are $f_0 = 0.97$, $f_1 = -1.65$, $f_2 = -1.14$, $f_3 = 3.86$, and $f_4 = -2.02$.

A representation of the power spectrum is given in Fig.~\ref{fig:power-Atmo-freq-cuts} with the exact boundary condition for the \Atmo \ model. As already emphasized in \citet{GB17}, the hypothesis of source equipartition leads to a power spectrum that compares well with the observations for frequencies below the cut-off. In particular, the peaks of high power are close to the observed eigenfrequencies with only a small shift due to the surface effect \citep{RC99}. For frequencies below the cut-off frequency, the power spectra obtained with the different boundary conditions are similar even though some differences are already visible by eye for Dirichlet. Above the cut-off frequency, the radiation boundary conditions and learned infinite elements perform well while of course the Dirichlet boundary condition is not adapted. More surprisingly at first sight, approximating the boundary condition around the cut-off frequency is the most challenging task. For $\ell > 500$, differences with the exact $\dtn$ are visible for low-order RBC and for learned infinite elements with $N=0$. High-order IE are necessary to emulate properly the wave behaviour around the cut-off which can be understood by the complex behaviour of the $\dtn$ as shown in Fig.~\ref{fig:dtn-approx-low-order}. These frequencies are interesting as they give information about the lower atmosphere \citep{Vorontsov1998}.

\subsection{Importance of the atmospheric model for the power spectrum and the eigenfrequencies}

\begin{figure}
\centering
\includegraphics[width=.5\textwidth]
{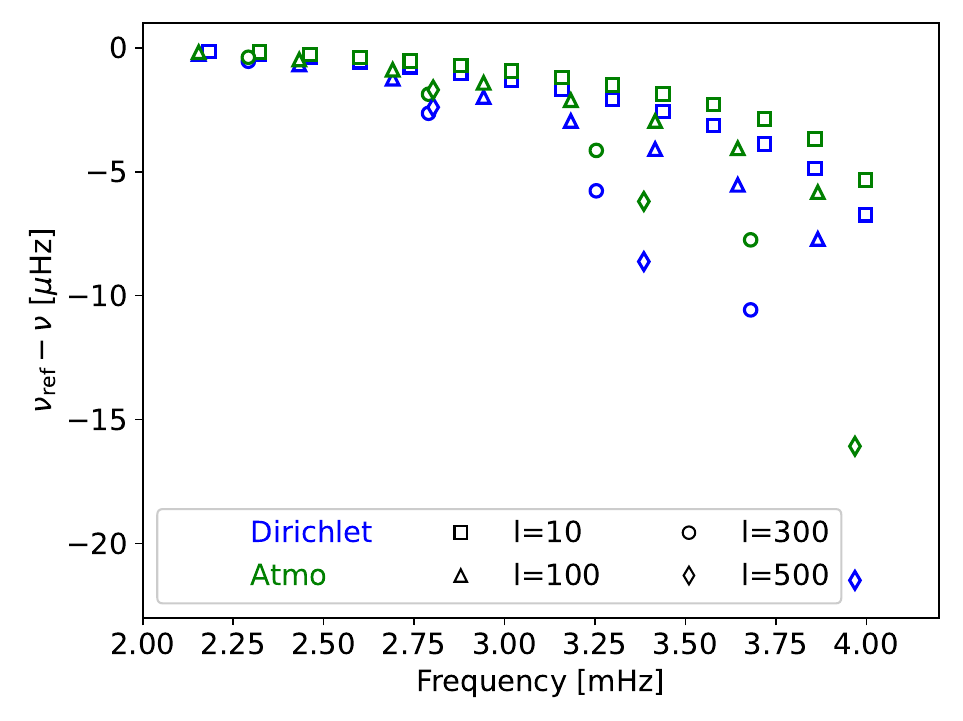}
\caption{Difference between the frequencies computed with the VAL-C atmospheric model (reference) and the \Atmo \ model (green) and Dirichlet boundary condition (blue) for different values of the harmonic degree $\ell$.}
\label{fig:error-freq}%
\end{figure}

To show the sensitivity of the eigenfrequencies to the atmosphere model, Fig.~\ref{fig:error-freq} represents the difference between the frequencies computed with the VAL-C model (considered as the reference) and the \Atmo \ model or a Dirichlet boundary condition. To do so, we measure the frequencies of the resonances by fitting Lorentzian functions to the peaks of the power spectrum for the different boundary conditions. The differences are important and way above the measurement error \citep[depending on the mode but of order 0.1~$\mu$Hz][]{Korzennik2013}.  This shows that the atmosphere plays a significant role in the surface effect even for frequencies way below the acoustic cut-off. To ensure that this conclusion is also valid for more sophisticated equations of solar oscillations, we computed the eigenfrequencies with GYRE \citep{TT13} using a Dirichlet and an isothermal atmosphere. The differences in the eigenfrequencies are similar to the ones reported in Fig.~\ref{fig:error-freq}.

\subsection{Accuracy of the learned infinite elements}

\begin{figure}
\centering
\includegraphics[width=.5\textwidth]{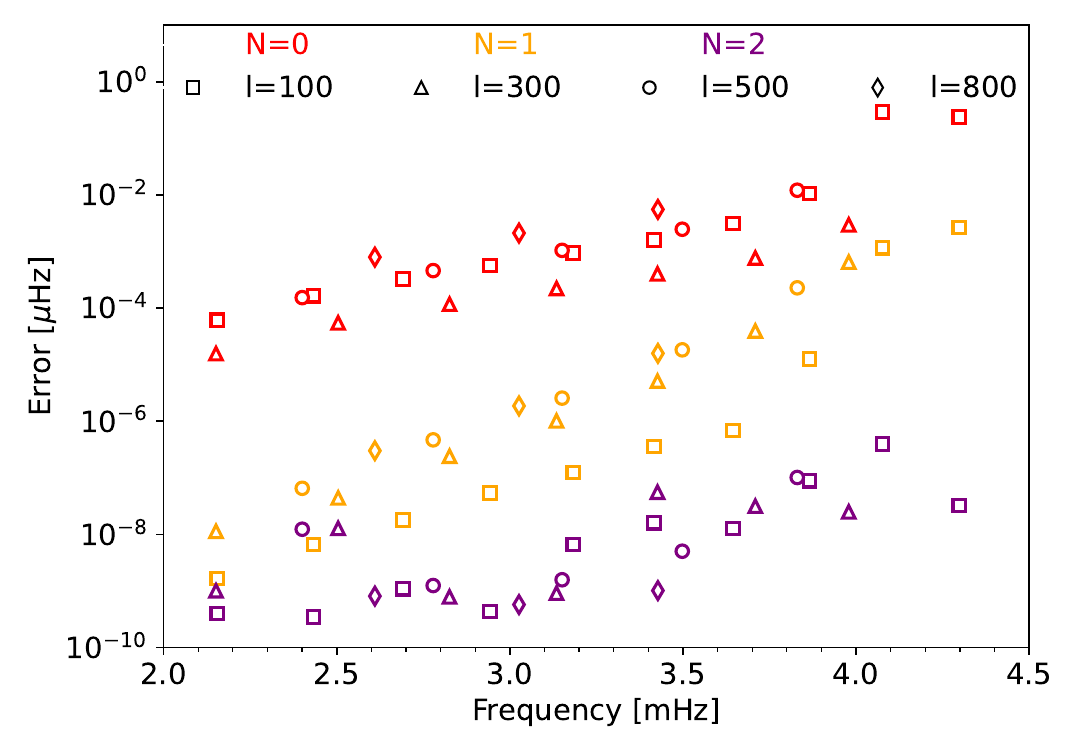}
\caption{Error in the location of the peaks of the modes for different values of the harmonic degree $\ell$ and for different boundary conditions for the VAL-C atmospheric model.}
\label{fig:error-val-freq}%
\end{figure}

In the previous section, we have shown that the atmosphere plays a significant role in the observed spectrum. The location of the eigenfrequencies was computed from the reference solution. In this section, we want to show that the VAL-C atmosphere can be emulated with sufficient accuracy using learned infinite elements. We refer to Appendix~\ref{section:accuracy_atmo} for a similar accuracy test for the \Atmo \ model. We compute the Green's function using learned infinite elements of different orders applied directly at the solar surface and compare with a reference obtained by applying the exact $\dtn$ from Eq.~\eqref{eq:dtn-ODE}. The error on the eigenfrequencies for the different levels of approximations is shown in Fig.~\ref{fig:error-val-freq}. The approximation of order 0 is sufficient only for frequencies smaller than 4~mHz and a condition of order 1 is necessary for higher frequencies to obtain an error below the measurement error.

\begin{figure*}
\centering
\includegraphics[width=\textwidth]{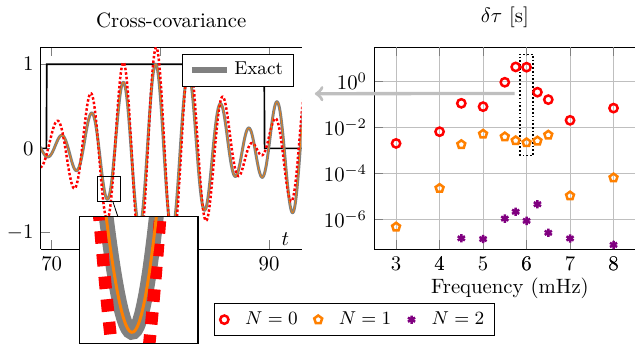}
	\caption{Left: Cross-covariance at $6.0$ mHz for the different boundary conditions and window function used to select the wave packet to compute travel times (black). Right: Travel time perturbation caused by the approximate boundary conditions for the VAL-C model across the frequency spectrum for two points separated by $\theta = 30 \degree$. }
\label{fig:delta-tau-VALC}%
\end{figure*}

For applications in time-distance helioseismology, we can also estimate the accuracy of the boundary condition in terms of travel times. To do so, we first need to define the expectation value of the cross-covariance function. It is obtained by summing up the harmonic coefficients of the power spectrum
\begin{equation}\label{eq:cross_covariance_modal}
	C(\theta,\omega) =  \sum\limits_{\ell} \frac{2 \ell +1}{4\pi} F_{\ell}(\omega) \mathcal{P}_\ell(\omega) P_{\ell}(\cos{\theta}),
\end{equation}
where $\theta$ is the angular distance between the two observation points that are cross-correlated. $P_\ell$ are the Legendre polynomials of degree $\ell$ and $F_\ell$ corresponds to a filter to select a certain range of frequencies or type of waves, e.g. a phase-speed filter. Here, we apply Gaussian frequency filters centered at different frequencies and with standard deviation $\sigma / 2\pi = 0.6$~mHz in order to study the error in different frequency bands.
Similar filters have been applied to compute frequency-dependent travel times  for meridional flow measurements \citep{RA20} even though the interpretation remains difficult due to systematic effects \citep{CZ18}. From the filtered cross-covariance, travel times between the reference and the approximate cross-covariances are computed using the linear formula from \citet{GB02}.


Figure~\ref{fig:delta-tau-VALC} shows the travel-time difference between the reference solution and approximate boundary conditions using learned infinite elements of different orders. For frequencies below 4~mHz, the learned IE with $N=0$ are sufficient to bring the error around 0.01~s which is smaller than the observation noise \citep{Liang2017}. Above 4~mHz and in particular around the cut-off frequency, learned IE of order 1 should be used to guarantee an error smaller than 0.01~s. Using higher-order (N=2) reduces drastically the error (around $10^{-6}$~s). The behaviour of the error is similar for the \Atmo \ model  (see Fig.~\ref{fig:delta-tau-Atmo}). However in this case, the learned infinite elements with $N=0$ are always sufficient to keep the error below 0.01~s.

\subsection{Computational cost}\label{ssection:comp-cost}

The previous section has shown that the accuracy provided by learned IEs placed at the surface $r_a = R_{\odot}$ using at most $N=2$ is already sufficient for incorporating the response from the complicated VAL-C atmosphere into the simulations. In this section we discuss the incurred computational costs and provide a comparison with a traditional approach in which the mesh is extended to include the entire atmosphere.
To this end, we consider a conforming finite element discretization of  Eq.~\eqref{eq:scalar_wave} in three dimensions ($3$D) or in an axisymmetric setting ($2.5$D, see \citet{GB17} for details) up to the surface $r_a = R_{\odot}$ using $10$ \dofs~per wavelength as reference.
Table~\ref{tab:comp-cost} shows the relative increase of \dofs~when adding learned infinite elements of degree $N$ to this discretization. For comparison, the last two columns give the increase when the mesh is extended to the end of Model S ($+0.5$ Mm above the surface) and to the end of the VAL-C model ($+2.5$ Mm), respectively.
\begin{table}[htbp]
 \label{tab:comp-cost}
 \caption{ Relative increase of \dofs~when a finite element discretization of Eq.~\eqref{eq:scalar_wave} is augmented by learned IEs of degree $N$. A naive approach of extending the volumetric mesh to $+0.5$ Mm or $+2.5 $ Mm above the surface is given in the last two columns.
  }
\begin{center}
\begin{tabular}{ccccccc} \toprule \hline
 & $N=0$ & $N\!=\!1$ & $N\!=\!2$   & $+0.5$ Mm  & $+2.5 $ Mm  \\ \hline
$2.5$D &  $0.0$ \%   & $4.3$ \%  &  $8.5$\%  &  $19.6$ \% & $72.9 $ \%   \\
$3$D  & $0.0$ \%  & $6.9 $ \%  &  $13.8 $ \%  & $59.3 $ \% & $203.4 $ \% \\
\bottomrule
\end{tabular}
\end{center}
\end{table}
The table clearly demonstrates that using learned IEs is significantly more efficient than meshing the entire atmosphere. The latter would increase the number of \dofs~by over $200$\% in three dimensions whereas learned IEs provide sufficient accuracy for an increase smaller than $14$\%. Moreover, the increase is already significant (around $60$\% ) when extending the mesh until the end of Model S ($+0.5$ Mm), a necessity to apply the radiation boundary conditions from \citet{BC18} for the \texttt{Atmo} model. Thus, applying learned IEs with say $N=1$ directly at the surface is clearly preferable as it only leads to an increase of around $7$\%.

\section{Conclusions and outlook} \label{section:conclusion}

Helioseismology often relies on the solution of a wave equation in the solar interior with simplified models of atmosphere (free surface or isothermal). We showed that the choice of the atmospheric model has repercussion on the position of the eigenvalues and generates artificial travel-time shifts even for waves at frequencies way below the acoustic cut-off. However, extending the computational domain to take into account the atmosphere leads to a significant increase of the computational cost. For example, the number of degrees of freedom is multiplied by 3 for 3D computations which renders them unfeasible. We propose to circumvent this problem by using learned infinite elements which emulate the behaviour of a 1D atmosphere without having to extend the computational domain. This approach is accurate and computationally efficient allowing to place the computational boundary directly at the observation height. Placing the boundary condition at the solar surface instead of 500~km above leads to a decrease of around $50$\% of the degrees of freedom for 3D computations, a significant improvement in this computationally challenging task of simulating wave propagation in a 3D medium with realistic solar stratification.

The high flexibility of learned IE renders them a suitable candidate to be employed in inversions for the solar atmosphere. The exact $\dtn$ in the minimization problem (Eq.~\eqref{eq:NLLSE}) could be replaced by observational constraints such as the solar power spectrum. By doing so, the synthetic power spectrum generated by solving Eq.~\eqref{eq:scalar_wave} would be as close as possible to the observed one. We have shown that a general model of atmosphere such as VAL-C could be emulated with only a few degrees of freedom and thus the inversion would have to recover only a few parameters making this method very attractive.

We did not consider background flows in this paper. The method presented here can be applied if flows are confined to the solar interior (e.g. meridional circulation) but they cannot be present in the atmosphere as it breaks the separability of the horizontal and radial variables, an essential property in the current design of learned infinite elements. This is an important limitation as it prevents us from treating rotation. First steps have been taken in \citet[][chapter 8]{JP_PHD_2021} for axisymmetric medium considering only an azimuthal order $m=0$ but additional work in this direction is necessary.

In this paper, we applied the learned infinite elements to a simplified scalar equation representative of the propagation of p-modes in the solar interior.
The next step is the extension to a more general form of the equation of stellar oscillations as introduced by \citet{LO67}. The well-posedness of this equation has been proven by \citet{HH20} and recent works deal with appropriate discretizations of this problem \citep{CD18,AHLS_ARXIV_2022,HLS_ARXIV_2022}. These efforts should be combined with a suitable transparent boundary condition. Radiation boundary conditions for an isothermal atmosphere have been proposed by \citet{Barucq2021} by transforming the vector problem into a scalar equation. The general case of a spherically symmetric atmosphere has been considered in \citet{MH21}. Therein, using the Cowling approximation a scalar equation in the atmosphere has been derived and coupled to the vector equation in the interior. Such a coupling paves the way for using learned infinite elements on the equations of \citet{LO67}.

\begin{acknowledgements}
      The authors acknowledge funding by the Deutsche Forschungsgemeinschaft (DFG, German Research Foundation) -- Project-ID 432680300 -- SFB 1456 (project C04).  Additionally, J.P. acknowledges funding by EPSRC grant EP/V050400/1.
\end{acknowledgements}

%
\bibliographystyle{aa} 
\bibliography{references} 
%


\begin{appendix}
\section{Implementation in context of FEM}\label{section:FEM-impl}
We sketch the implementation of the boundary conditions covered in this article for the FEM. We refer to \citet{HLP21,JP_PHD_2021} for more details . Let us suppose that the discretization of the first term in Eq.~\eqref{eq:scalar_wave_weak}, i.e.\ excluding terms on $\Gamma$, leads to a linear system of the form
\begin{equation}\label{eq:interior-matrix}
\begin{bmatrix} L_{II} & L_{I\Gamma}\\ L_{\Gamma I}& L_{\Gamma\Gamma}^{\mathrm{int}} \end{bmatrix}
\begin{bmatrix}\psih_I \\ \psih_\Gamma\end{bmatrix}
= \begin{bmatrix}\sh_I \\ \sh_\Gamma^{\mathrm{int}}\end{bmatrix}.
\end{equation}
Here, we partitioned the \dofs~as $\psih_{I}$ belonging to the
interior of the domain and a vector $\psih_\Gamma$ of \dofs~on $\Gamma$.
\subsection{The ideal boundary condition}
Let $\FEshape_{i}$ for $i=1,\ldots,n_{\Gamma}$ denote the basis functions of the FEM with support on $\Gamma$.
Further, let $\massFEM$ and $\stiffFEM$ denote mass and stiffness matrices of the FEM, i.e.\
\begin{equation}\label{eq:def_massFEM}
\massFEM_{i j} = \int_{ \Gamma }{ \frac{1}{\rho} \FEshape_{j}  \FEshape_{i} \, \mathrm{d}S   }, \quad
\stiffFEM_{i j} = \int_{ \Gamma }{ \frac{1}{\rho} \nabla_{\Gamma} \FEshape_{j}  \nabla_{\Gamma} \FEshape_{i} \, \mathrm{d}S.   }
\end{equation}
Ideally, one would add a matrix representation $ \massFEM \dDtN^{\mathrm{ext}} $ of the second term  in Eq.~\eqref{eq:scalar_wave_weak} to ~\eqref{eq:interior-matrix}, which implements the exact boundary conditions given in ~\eqref{eq:exact-Dtn-coupling-exterior}.
This would lead to a linear system of the form
\begin{equation}\label{eq:matrix-formu-w-exact-DtN}
\begin{bmatrix} L_{II} & L_{I\Gamma}\\ L_{\Gamma I}& L_{\Gamma\Gamma}^{\mathrm{int}} + \massFEM \dDtN^{\mathrm{ext}} \end{bmatrix}
\begin{bmatrix}\psih_I \\ \psih_\Gamma\end{bmatrix}
= \begin{bmatrix}\sh_I \\ \sh_\Gamma^{\mathrm{int}}\end{bmatrix}.
\end{equation}
This approach is, however, not practical as it leads to many issues.
Foremost, $\dDtN^{\mathrm{ext}}$ will be a dense matrix
which is computationally very expensive to assemble and
incorporate into the otherwise sparse linear system of the FEM.

\subsection{Affine approximation}
According to Eq.~\eqref{eq:impl-low-order-FEM}, the affine approximation $ \DtN^{0}$ of $\DtN$ leads to a linear system of the form
\begin{equation}\label{eq:matrix-formu-DtN-lin}
\begin{bmatrix} L_{II} & L_{I\Gamma}\\ L_{\Gamma I}& L_{\Gamma\Gamma}^{\mathrm{int}} + L_{\Gamma\Gamma} \end{bmatrix}
\begin{bmatrix}\psih_I \\ \psih_\Gamma\end{bmatrix}
= \begin{bmatrix}\sh_I \\ \sh_\Gamma^{\mathrm{int}}\end{bmatrix}
\end{equation}
for $L_{\Gamma\Gamma} = \alpha \massFEM  + \beta \stiffFEM $. Comparing with ~\eqref{eq:matrix-formu-w-exact-DtN}, this corresponds to the approximation $L_{\Gamma\Gamma} \approx \massFEM \dDtN^{\mathrm{ext}}$.

\subsection{Rational approximation}
As described in Sect.~\ref{section:infel} and Fig.~\ref{fig:learnedIE-sketch-Sun}, we implement the rational approximation $\DtN^{N}$ of $\DtN$ by introducing additional \dofs~$\psih_E$ in the exterior. The linear system to be solved is then of the form
\begin{equation}\label{eq:matrix-formu-DtN-rat}
\begin{bmatrix} L_{II} & L_{I\Gamma} & 0 \\
L_{\Gamma I} & L_{\Gamma\Gamma}^{\mathrm{int}} + L_{\Gamma\Gamma} & L_{\Gamma E}  \\
0 & L_{E \Gamma } & L_{E E}
\end{bmatrix}
\begin{bmatrix}\psih_I \\ \psih_\Gamma \\ \psih_E \end{bmatrix}
= \begin{bmatrix}\sh_I \\ \sh_\Gamma^{\mathrm{int}} \\ 0\end{bmatrix}
\end{equation}
where
\begin{equation}\label{eq:ansatz}
\begin{bmatrix}  L_{\Gamma\Gamma} & L_{\Gamma E} \\ L_{E\Gamma} & L_{EE}\end{bmatrix}
= \dstiffr \otimes \massFEM + \dmassr\otimes \stiffFEM.
\end{equation}
Here, $\otimes $ denotes the Kronecker product of matrices and
\begin{equation*}
\dstiffr =
 \begin{bmatrix}
         \dstiffr_{00}  & \dstiffr_{01}  &   \cdots  &  \cdots &  \dstiffr_{0  N }  \\
    \dstiffr_{01}     &   \ddots       &   &  \mathbf{0} &  \\
    \vdots &  & \ddots   &  &  \\
        \vdots & \mathbf{0}  &   & \ddots  &  \\
    \dstiffr_{ 0 N}  &  &     &  & \dstiffr_{ N N}    \\
    \end{bmatrix},
  \;
  \dmassr =
 \begin{bmatrix}
   \dmassr_{00}  & \dmassr_{01} &   \cdots  &  \cdots &  \dmassr_{0 N} \\
     \dmassr_{01}      &   1     &   &  \mathbf{0} &  \\
    \vdots &  & \ddots   &  &  \\
        \vdots & \mathbf{0}  &   & \ddots  &  \\
   \dmassr_{0N}   &  &     &  & 1   \\
    \end{bmatrix},
     \end{equation*}
where the non-zero matrix entries are obtained from solving the minimization problem given in Eq.~\eqref{eq:NLLSE}. By eliminating the exterior \dofs~$\psih_E$, it is possible to rewrite the linear system of Eq.~\eqref{eq:matrix-formu-DtN-rat} in the compressed form
\begin{equation}\label{eq:matrix-formu-DtN-rat-compressed}
\begin{bmatrix} L_{II} & L_{I\Gamma}\\ L_{\Gamma I}& L_{\Gamma\Gamma}^{\mathrm{int}} + \massFEM \dDtN^N  \end{bmatrix}
\begin{bmatrix}\psih_I \\ \psih_\Gamma\end{bmatrix}
= \begin{bmatrix}\sh_I \\ \sh_\Gamma^{\mathrm{int}}\end{bmatrix}
\end{equation}
for
\begin{equation}\label{eq:dDtN}
\dDtN^N := \massFEM^{-1}\left(L_{\Gamma \Gamma} - L_{\Gamma E}L_{EE}^{-1}L_{E\Gamma}\right),
\quad \dDtN^N \approx \dDtN^{\mathrm{ext}}.
\end{equation}
Comparing with the affine  approximation, we see that the term $L_{\Gamma\Gamma}$ approximating  $\massFEM \dDtN^{\mathrm{ext}} $ has been enriched by the Schur complement with respect to the exterior \dofs. Note that in practice, it is preferable to solve the sparse system~\eqref{eq:matrix-formu-DtN-rat} rather than~\eqref{eq:matrix-formu-DtN-rat-compressed}  which contains the dense block $\massFEM \dDtN_N$ on the interface $\Gamma$.

Finally, let us explain why Eq.~\eqref{eq:dDtN} corresponds to a rational approximation of $\DtN$. In analogy with Eq.~\eqref{eq:spherical-harmonics-eigenfunctions}, let $\underline{Y}_{\ell}$ denote the discrete eigenfunctions of the horizontal Laplacian, which fulfill $\stiffFEM \underline{Y}_{\ell} = \underline{\lambda}_{\ell} \massFEM \underline{Y}_{\ell}$ for discrete eigenvalues $\underline{\lambda}_{\ell} \approx \lambda_{\ell}$. Then it has been shown in \citep[Proposition 2.1]{HLP21} that
\begin{equation}\label{eq:DtN-N-discrete}
    \dDtN^N \underline{Y}_\ell = \dtn^{N}_{\ell} \underline{Y}_\ell,
\end{equation}
where $\dtn^{N}_{\ell}$ is the rational function from Eq.~\eqref{eq:dtn-N}. That is, $\dDtN^N$ emulates the rational $ \DtN^{N}$ operator of Eq.~\eqref{eq:DtN-N-discrete} at the discrete level.

\section{Optimal weights of the learned infinite elements for $N=0$} \label{section:minimization}

The least-square problem to obtain the optimal coefficients $\alpha$ and $\beta$ for the learned IE of order 0 can be solved analytically. The pseudo-inverse $\Lambda^\dagger$ is defined as
\begin{equation}
    \Lambda^\dagger = (\Lambda^T \Lambda)^{-1} \Lambda^T b.
\end{equation}
The matrix $\Lambda^T \Lambda$ can be computed and inverted analytically and we obtain
\begin{align}
\alpha &= \frac{1}{| \Lambda^T \Lambda | }  \left[ \sum_{\ell=1}^L w_\ell^2 \lambda_\ell^2  \sum_{\ell=1}^L w_\ell^2 \dtn_\ell  -  \sum_{\ell=1}^L w_\ell^2 \lambda_\ell  \sum_{\ell=1}^L w_\ell^2 \lambda_\ell \dtn_\ell \right], \\
\beta &= \frac{1}{| \Lambda^T \Lambda | }  \left[ \sum_{\ell=1}^L w_\ell^2 \lambda_\ell \sum_{\ell=1}^L w_\ell^2 \dtn_\ell -  \sum_{\ell=1}^L w_\ell^2  \sum_{\ell=1}^L w_\ell^2 \lambda_\ell \dtn_\ell  \right],
\end{align}
where
\begin{equation}
    | \Lambda^T \Lambda | = \left( \sum_{\ell=1}^L w_\ell^2 \right) \left( \sum_{\ell=1}^L w_\ell^2 \lambda_\ell^2 \right) - \left( \sum_{\ell=1}^L w_\ell^2 \lambda_\ell \right)^2.
\end{equation}

\section{Accuracy of the boundary conditions for the \Atmo \ model} \label{section:accuracy_atmo}
\begin{figure*}
\centering
\includegraphics[width=\textwidth]{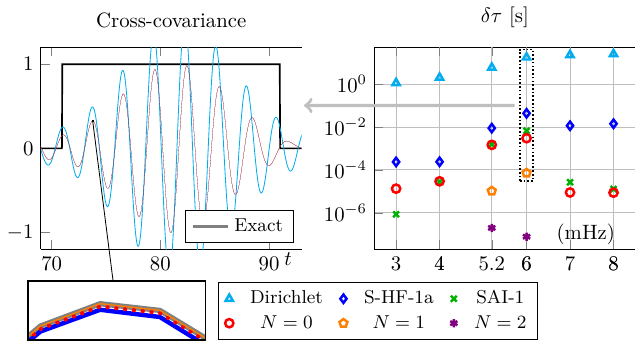}
	\caption{Left: cross-covariance obtained after applying a frequency filter around $6.0$ mHz for different approximate boundary conditions of the \texttt{Atmo} model. Right: associated travel time perturbation using frequency filters centered at different frequencies for a separation distance $\theta = 30 \degree$ between the two observation points (the perturbations for the $N=1$ and $N=2$ conditions are smaller than $10^{-6}$s for frequencies below $4$ mHz and higher than $6$ mHz and therefore not visible in the plot).}
\label{fig:delta-tau-Atmo}%
\end{figure*}

\begin{figure}
\centering
\includegraphics[width=.5\textwidth]{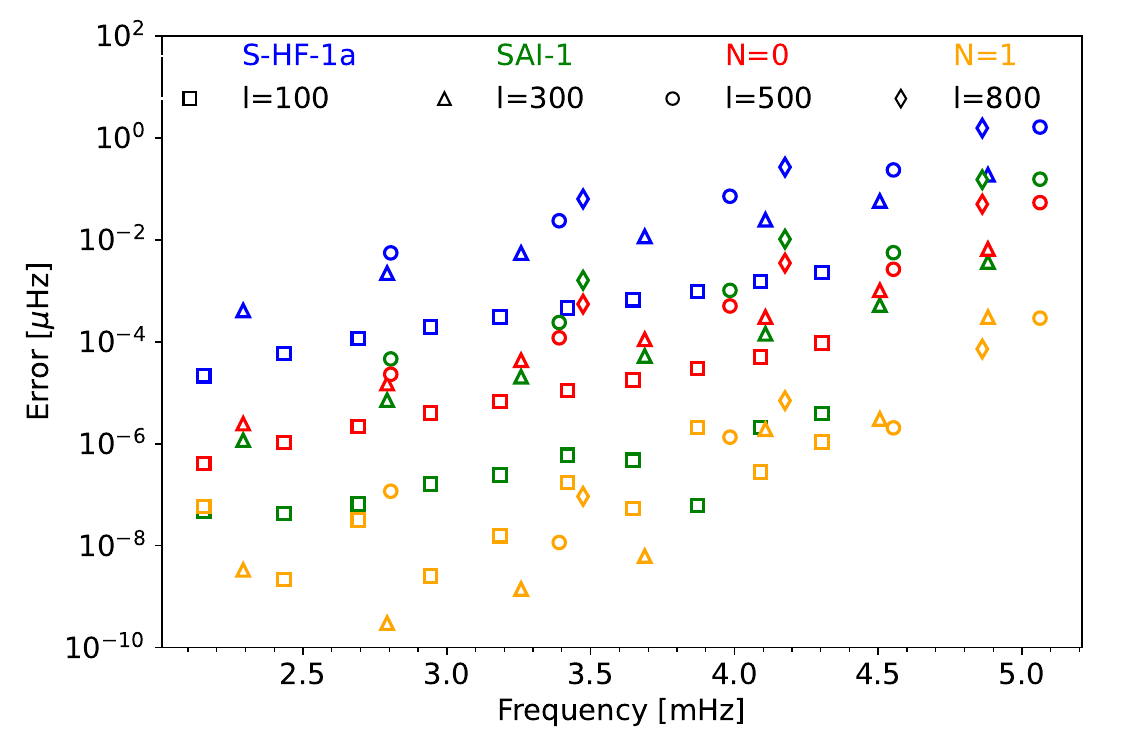}
\caption{Error in the location of the peaks of the modes for different values of the harmonic degree $\ell$ and for different boundary conditions for the \texttt{Atmo} model.}
\label{fig:error-atmo-freq}%
\end{figure}

This appendix presents the error in terms of location of the eigenfrequencies and travel times using approximate boundary conditions for the \Atmo \ model. Figures~\ref{fig:delta-tau-Atmo}~and~\ref{fig:error-atmo-freq} are the equivalent of Figs.~\ref{fig:delta-tau-VALC}~and~\ref{fig:error-freq} for the \Atmo \ model instead of VAL-C.

Figure~\ref{fig:error-atmo-freq} shows the error on the eigenvalues between an approximate boundary condition and the exact $\dtn$. The frequencies of the resonances are measured by fitting a Lorentzian to the power spectrum for the different boundary conditions. For all the boundary conditions the error is increasing with frequency and harmonic degree. Contrary to the VAL-C model (Fig.~\ref{fig:error-freq}), the low-order radiation boundary condition SAI-1 or the learned IE of order 0 already give errors below the measurement uncertainty.

The error in terms of travel times is shown in Fig.~\ref{fig:delta-tau-Atmo}. Travel times in different ranges of frequencies are obtained by applying a Gaussian filter with standard deviation $\sigma / 2\pi = 0.6$~mHz. The conclusions are similar to the study of the error in the eigenvalues, that is, the SAI-1 boundary condition or the learned infinite elements of order 0 are sufficient to obtain an error smaller than 0.01~s which is below the measurement error for e.g. meridional circulation measurements \citep{Liang2017}.

\end{appendix}

\end{document}